\documentclass[11pt]{revtex4}
\usepackage{amsmath}
\usepackage{amssymb,epsf}
\usepackage{epsfig}
\usepackage{epstopdf}
\usepackage{graphicx}
\usepackage{color}

\begin{document}

\title{Extended phase space thermodynamics and \\
$P$--$V$ criticality of black holes with a nonlinear source}
\author{S. H. Hendi$^{1,2}$\footnote{email address: hendi@shirazu.ac.ir} and M. H. Vahidinia$^{1,3}$\footnote{
email address: vahidinia@shirazu.ac.ir}}
\affiliation{$^1$Physics Department and Biruni Observatory, College of Sciences, Shiraz
University, Shiraz 71454, Iran \\
$^2$Research Institute for Astrophysics and Astronomy of Maragha (RIAAM),
P.O. Box 55134-441, Maragha, Iran\\
$^3$Physics Department, College of Sciences, Salman Farsi University Of
Kazerun, Kazerun, Iran }

\begin{abstract}
In this paper, we consider the solutions of Einstein gravity in
the presence of a generalized Maxwell theory, namely power Maxwell
invariant. First, we investigate the analogy of nonlinear charged
black hole solutions with the Van der Waals liquid--gas system in
the extended phase space where the cosmological constant appear as
pressure. Then, we plot isotherm $P$--$V$ diagram and study the
thermodynamics of AdS black hole in the (grand canonical)
canonical ensemble in which (potential) charge is fixed at
infinity. Interestingly, we find the phase transition occurs in
the both of canonical and grand canonical ensembles in contrast to
RN black hole in Maxwell theory which only admits canonical
ensemble phase transition. Moreover, we calculate the critical
exponents and find their values are the same as those in mean
field theory. Besides, we find in the grand canonical ensembles
universal ratio $\frac{P_{c}v_{c}}{T_{c}}$ is independent of
spacetime dimensions.
\end{abstract}

\maketitle

%%%%%%%%%%%%%%%%%%%%%%%%%%%%%%%%%%%%%

\section{Introduction}

Theoretically one may be expect the cosmological constant term to
arise from the vacuum expectation value of a quantum field and
hence can vary. Therefore, it may be considered in the first law
of thermodynamics with its conjugate
\cite{Gibbons1,BrownPLB1987,CaldarelliCQG2000}. By this
generalization, the cosmological constant and its conjugate can be
interpreted as geometrical pressure and volume of a black object
system, respectively. Moreover, this approach leads to an
interesting conjecture on reverse isoperimetric inequality for
black holes in contrast to a Euclidean version of isoperimetric
inequality. Regarding the inequality conjecture, some of the black
hole processes may be restricted
\cite{RayCQG2009,Gibbons2,PVpapers}.

Furthermore, the extension of thermodynamic phase space has dramatic effects
on the studying of famous phase transition of black holes in AdS space \cite%
{MyersPRD1999,Banerjee:2011au,Wu2012} and improves the analogy between
small/large black hole with the Van der Waals liquid/gas phase transitions.
Indeed, the AdS charged black holes exhibit an interesting phase transition
with the same critical behavior as Van der Waals model, qualitatively \cite%
{PVpapers}.

Taking into account the above statements, we should note that the
charge of the black hole plays a crucial role in this phase
transition. Therefore it is important to know effects of any
modification in the electromagnetic field. Indeed, some
characteristic features of universality class of phase
transitions such as the value of critical exponents or universal ratio $\frac{%
P_{c}v_{c}}{T_{c}}$ may depend on electromagnetic source or
spacetime dimension. On the other hand, considering strong
electromagnetic field in regions near to point-like charges, Dirac
suggested that one may have to use generalized nonlinear Maxwell
theory in those regions \cite{Dirac}. Similar behavior may occur
in the vicinity of neutron stars and black objects and so it is
expected to consider nonlinear electromagnetic fields with an
astrophysical motive \cite{Bialynicka}. In addition, within the
framework of quantum electrodynamics, it was shown that quantum
corrections lead to nonlinear properties of vacuum which affect
the photon propagation
\cite{Heisenberg,Delphenich,Schwinger,Stehle}. Moreover, the
effects of Born--Infeld (BI) source in the thermodynamics and
phase transition of black hole \cite{ThermoBI} have been studied.
Besides, in context AdS/CFT some authors consider roles of BI
source on shear viscosity \cite{Sun08} and holographic
superconductors \cite{AdSCFTBI}.

By this observations one may find it is worthwhile to study the
effects of nonlinear electrodynamics (NLEDs) on phase transition
of black holes in the extended phase space. In this direction, the
effects of nonlinear electromagnetic field of static and rotating
AdS black holes in the extended phase space have been analyzed
\cite{PVnonlinear}. It has been shown that for the BI black holes,
one may obtain the same qualitative behavior as RN black holes.
Indeed, BI electromagnetic field does not have any effect on the
values of critical exponents, but it changes the universal ratio $\rho_{c}=%
\frac{P_{c}v_{c}}{T_{c}}$ \cite{PVnonlinear}.

Although BI theory is a specific model in the context of NLEDs,
the recent interest on the NLEDs theories is mainly due to their
emergence in the context of low-energy limit of heterotic string
theory or as an effective action for the consideration of effects
loop corrections in QED where quartic corrections of Maxwell field
strength appear \cite{Kats}.

In the last five years, a class of NLEDs has been introduced, the
so-called power Maxwell invariant (PMI) field (for more details,
see \cite{PMIpapers1,PMIpapers2}). The PMI field is significantly
richer than that of the Maxwell field, and in the special case
($s=1$) it reduces to linear electromagnetic source. The black
hole solutions of the Einstein-PMI theory and their interesting
thermodynamics and geometric properties have been examined before
\cite{PMIpapers1,PMIpapers2}. In addition, in the context of
AdS/CFT correspondence, the effects of PMI source on strongly
coupled dual gauge theory have been investigated \cite{AdSCFTPMI}.

The bulk action of Einstein-PMI gravity has the following form \cite%
{PMIpapers2}
\begin{equation}
I_{b}=-\frac{1}{16\pi }\int_{M}d^{n+1}x\sqrt{-g}\left( R+\frac{n(n-1)}{l^{2}}%
+\mathcal{L}_{PMI}\right) ,  \label{Action}
\end{equation}%
where $\mathcal{L}_{PMI}=(-\mathcal{F})^{s}$ and $\mathcal{F}=F_{\mu \nu
}F^{\mu \nu }$. Before we proceed, we provide some of reasonable motivation
for considering this form of NLEDs.

\emph{First}, between NLEDs theories, the PMI theory is a toy model to
generalize Maxwell theory which reduces to it for $s=1$. One of the most
important properties of the PMI model in $(n+1)$-dimensions occurs for $%
s=(n+1)/4$ where the PMI theory becomes conformally invariant and
so the trace of energy-momentum tensor vanishes, the same as
Maxwell theory in four-dimensions \cite{PMIpapers1}. Considering
this value for the nonlinearity parameter, $s$, one can obtain
inverse square law for the electric field of charged pointlike
objects in arbitrary dimensions (the same as Coulomb's field in
four-dimensions). Furthermore, it has been shown that there is an
interesting relation between the solutions of a class of
pure $F(R)$ gravity and those of conformally invariant Maxwell source ($%
s=(n+1)/4$) in Einstein gravity \cite{CIMFR}.

\emph{Second}, we should note that considering the $E_{8}\times
E_{8}$ heterotic string theory, the $SO(32)$ gauge group has a
$U(1)$ subgroup. It has been shown that \cite{StrinNL} taking into
account a constant dilaton, the effective Lagrangian has
Gauss-Bonnet term as well as a quadratic Maxwell invariant in
addition to the Einstein-Maxwell Lagrangian. Since, unlike the
quadratic Maxwell invariant, the Gauss-Bonnet term becomes a
topological invariant and does not give any contribution in
four-dimensions, it is natural to investigate Einstein-NLEDs in
four dimensions. Taking into account a PMI theory as a NLEDs
Lagrangian and expanding it for $\mathcal{F\longrightarrow F}_{0}$
(where we considered $\mathcal{F}_{0}$ as a unknown constant which
we should fix it.), we find
\begin{equation}
\mathcal{L}_{PMI}\simeq -a_{1}\mathcal{F}+(s-1)\left[ a_{0}+a_{2}(-\mathcal{F%
})^{2}+a_{3}(-\mathcal{F})^{3}+...\right] .  \label{Expand}
\end{equation}%
In other words, one can consider series expansion of $\mathcal{L}_{PMI}$
near a constant $\mathcal{F}_{0}$ and obtain Eq. (\ref%
{Expand}), in which the constants $a_{i}$'s are depend on $s$ and $\mathcal{F%
}_{0}$. In order to obtain $\mathcal{F}_{0}$ and also have a
consistent series expansion with linear Maxwell Lagrangian
($s=1$), one should set $a_{1}=1$. Taking into account $a_{1}=1$
and obtaining $\mathcal{F}_{0}$, we are in a position to get a new
series expansion for $\mathcal{L}_{PMI}$
\begin{equation}
\mathcal{L}_{PMI}\simeq -\mathcal{F}+(s-1)\left[ b_{0}+b_{2}(-\mathcal{F}%
)^{2}+b_{3}(-\mathcal{F})^{3}+...\right] ,  \label{Expand2}
\end{equation}%
where $b_{i}$'s are only depend on $s$. Although, $\mathcal{L}_{PMI}=(-%
\mathcal{F})^{s}$ can lead to Eq. (\ref{Expand2}) by a series expansion,
working with Eq. (\ref{Expand2}) is more complicated and we postpone the
study of this scenario to another paper.

\emph{Third}, taking into account the applications of the AdS/CFT
correspondence to superconductivity, it has been shown that the PMI theory
makes crucial effects on the condensation as well as the critical
temperature of the superconductor and its energy gap \cite{AdSCFTPMI}.

Motivated by the recent results mentioned above, we consider the
PMI theory to investigate the effects of nonlinearity on the
extended phase space thermodynamics and $P$--$V$ criticality of
the solutions. Moreover, to better understand the role of
nonlinearity, we relax the conformally invariant constraint and
take $s$ as an arbitrary constant. It helps us to have a deep
perspective to study the universal behavior of large/small black
hole phase transitions. In particular, we are keen on
understanding sensitivity of the critical exponents, universal
ratio and other thermodynamic properties to nonlinearity
parameter, $s$.

Outline of this paper is as follows: In Sec. \ref{BH}, we consider
spherically symmetric black hole solutions of Einstein gravity in
the presence of the PMI source. Regarding the cosmological
constant as thermodynamic pressure, we study thermodynamic
properties and obtain Smarr's mass relation. In Sec. \ref{Cano},
we investigate the analogy of black holes with Van der Waals
liquid--gas system in the grand canonical ensemble by fixing
charge at infinity. In this ensemble we find the free energy and
plot the coexistence curve of a small/large black hole. Then, we
calculate the critical exponent and find they match to mean field
value (same as a Van der Waals liquid). Moreover, we consider the
special case $s=n/2$ as BTZ-like solution, study its phase
transition and show the critical exponents are the same as former
case. In Sec. \ref{GCano}, we consider the possibility of the
phase transition in the grand canonical ensemble and find that in
contrast to RN black holes, the phase transition occurs for $s
\neq 1$. Finally, we finish this work with some concluding
remarks.

\section{Extended phase-space thermodynamics of black holes with PMI source}

\label{BH}

We consider a spherically symmetric spacetime as
\begin{equation}
ds^{2}=-f(r)dt^{2}+\frac{dr^{2}}{f(r)}+r^{2}d\Omega _{d-2}^{2},
\label{Metric}
\end{equation}
where $d\Omega _{d}^{2}$ stands for the standard element on $S^{d}$.
Considering the field equations following from the variation of the bulk
action with Eq. (\ref{Metric}), one can show that the metric function $f(r)$%
, gauge potential one--form $A$ and electromagnetic field two--form $F$ are
given by \cite{PMIpapers2}
\begin{eqnarray}
f(r) &=&1+\frac{r^{2}}{l^{2}}-\frac{m}{r^{n-2}}+\frac{(2s-1)^{2}\left( \frac{%
(n-1)(2s-n)^{2}q^{2}}{(n-2)(2s-1)^{2}}\right) ^{s}}{%
(n-1)(n-2s)r^{2(ns-3s+1)/(2s-1)}},  \label{metfunction} \\
A &=&-\sqrt{\frac{n-1}{2(n-2)}}qr^{(2s-n)/(2s-1)}dt,  \label{A} \\
F&=&dA.  \label{dA}
\end{eqnarray}
The power $s \neq n/2$ denotes the nonlinearity parameter of the source
which is restricted to $s>1/2$ \cite{PMIpapers2}, and the parameters $m$ and
$q$ are, respectively, related to the ADM mass $M$ and the electric charge $%
Q $ of the black hole
\begin{eqnarray}
M &=&\frac{\omega _{n-1}}{16\pi }(n-1)m,  \label{Mass} \\
Q &=&\frac{\sqrt{2} (2s-1)s\; \omega _{n-1}}{8\pi }\left( \frac{n-1}{n-2}%
\right) ^{s-1/2}\left( \frac{\left( n-2s\right) q}{2s-1}\right) ^{2s-1},
\label{Charge}
\end{eqnarray}
where $\omega _{n-1}$ is given by
\begin{equation}
\omega _{n-1}=\frac{2\pi ^{\frac{n}{2}}}{\Gamma \left( \frac{n}{2}\right) }.
\label{Omega}
\end{equation}
It has been shown that \cite{PMIpapers2} Eqs. (\ref{Metric}) and (\ref%
{metfunction}) describe a black hole with a cauchy horizon ($r_{-}$) and an
event horizon ($r_{+}$). The event horizon radius of this black hole can be
calculated numerically by finding the largest real positive root of $%
f(r=r_{+})=0$. Using the surface gravity relation, we can obtain the
temperature of the black hole solutions as
\begin{equation}
T=\frac{f^{\prime }(r_{+})}{4\pi }=\frac{n-2}{4\pi r_{+}}\left( 1+\frac{n}{%
n-2}\frac{r_{+}^{2}}{l^{2}}-\frac{(2s-1)\left( \frac{(n-1)(2s-n)^{2}q^{2}}{%
(n-2)(2s-1)^{2}}\right) ^{s}}{(n-1)(n-2)r_{+}^{2(ns-3s+1)/(2s-1)}}\right).
\label{T}
\end{equation}

The electric potential $\Phi $, measured at infinity with respect
to the horizon while the black hole entropy $S$, was determined
from the area law. It is easy to show that
\begin{eqnarray}
\Phi &=&\sqrt{\frac{n-1}{2(n-2)}}\frac{q}{r_{+}^{(n-2s)/(2s-1)}},
\label{Phi} \\
S &=&\frac{\omega _{n-1}r_{+}^{n-1}}{4}.  \label{S}
\end{eqnarray}
Now, as it was considered before \cite{PVpapers}, we interpret
$\Lambda $ as a thermodynamic pressure $P$,
\begin{equation}
P=-\frac{1}{8\pi }\Lambda =\frac{n(n-1)}{16\pi l^{2}},  \label{PLambda}
\end{equation}
where its corresponding conjugate quantity is the thermodynamic volume \cite%
{Gibbons2}
\begin{equation}
V=\frac{\omega _{n-1}{r_{+}}^{n}}{n}.  \label{volrp}
\end{equation}
Considering obtained quantities, one can show that they satisfy the
following Smarr formula
\begin{equation}
M=\frac{n-1}{n-2}TS+\frac{ns-3s+1}{s(2s-1)(n-2)}\Phi Q-\frac{2}{n-2}VP.
\label{Smarr}
\end{equation}
It has been shown that Eq. (\ref{Smarr}) may be obtained by a scaling
dimensional argument \cite{scaling,RayCQG2009}. In addition, the (extended
phase-space) first law of thermodynamics can be written as
\begin{equation}
dM=TdS+\Phi dQ+VdP.  \label{firstLaw}
\end{equation}
In what follows, we shall study the analogy of the liquid--gas phase
transition of the Van der Waals fluid with the phase transition in black
hole solutions in the presence of PMI source.

\section{Canonical ensemble}

\label{Cano}

In order to study the phase transition, one can select an ensemble
in which black hole charge is fixed at infinity. Considering the
fixed charge as an extensive parameter, the corresponding ensemble
is called a canonical ensemble.

\subsection{Equation of state}

\begin{figure}[tbp]
$%
\begin{array}{cc}
\epsfxsize=7cm \epsffile{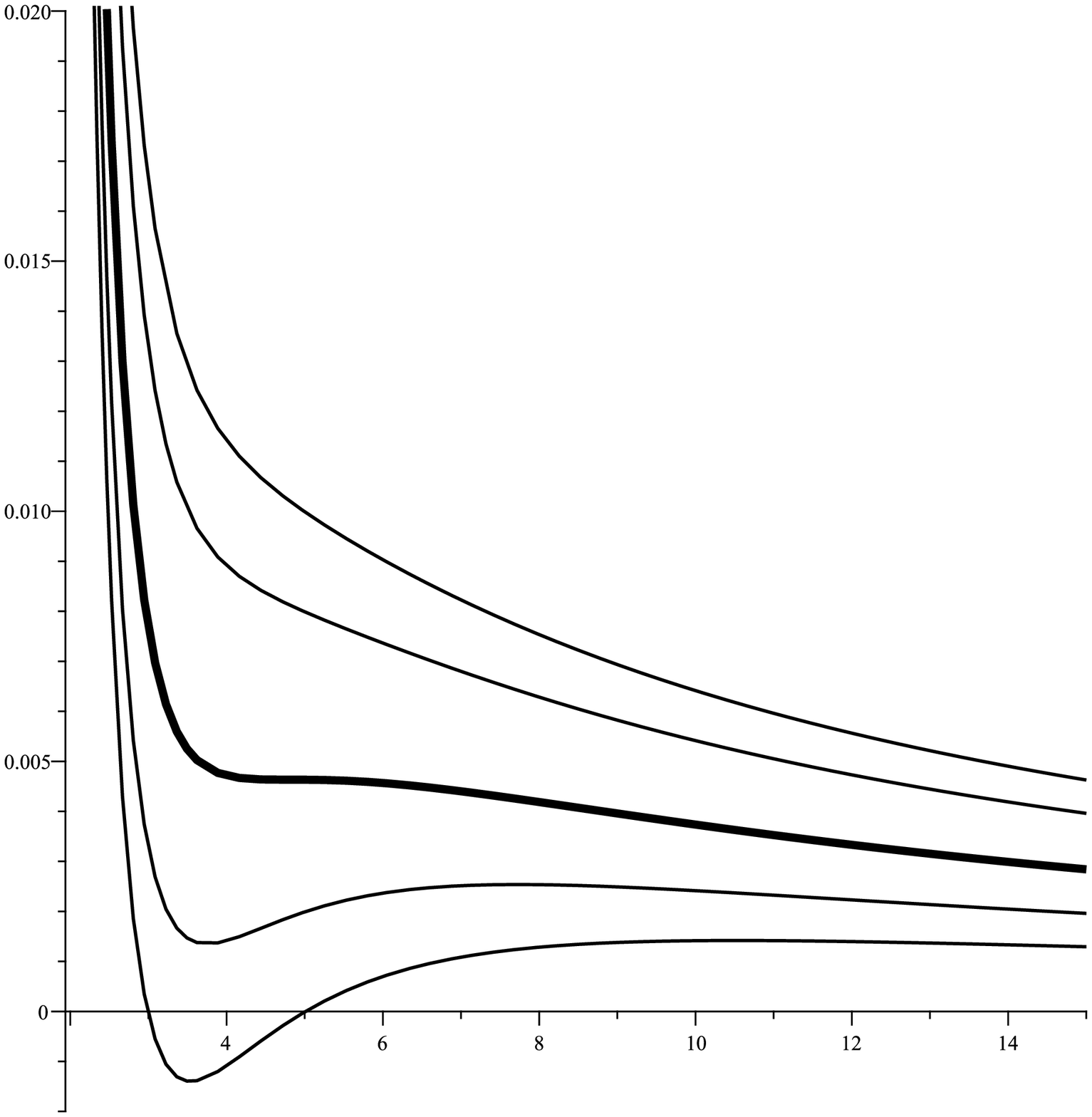} & \epsfxsize=7cm \epsffile{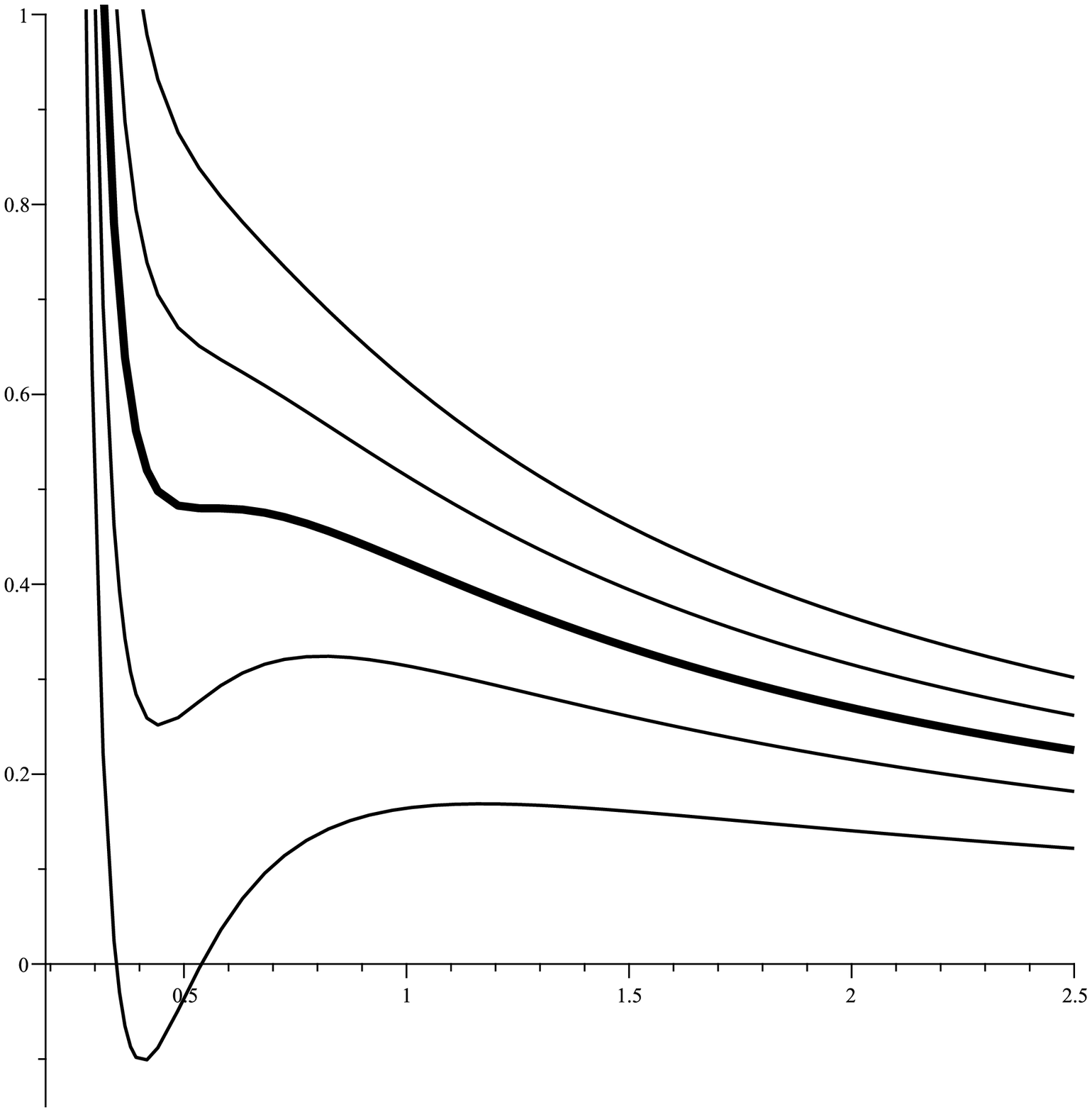}%
\end{array}
$%
\caption{$P-V$ diagram of charged AdS black holes in PMI for
$s=\frac{3}{4}$ with $n=3$ (left) and $s=2$ with $n=5$ (right).
The temperature of isotherms decreases from top to bottom. The
bold line is the critical isotherm diagram.} \label{FPV}
\end{figure}

Using the Eqs. (\ref{PLambda}) and (\ref{T}) for a fixed charge $Q$, one may
obtain the equation of state, $P(V,T)$
\begin{equation}
P=\frac{(n-1)}{4r_{+}}T-\frac{(n-1)(n-2)}{16\pi r_{+}^{2}}+\frac{1}{16\pi }%
\frac{(2s-1)\left( \frac{(n-1)(2s-n)^{2}q^{2}}{(n-2)(2s-1)^{2}}\right) ^{s}}{%
r_{+}^{2s(n-1)/(2s-1)}},  \label{state}
\end{equation}%
where $r_{+}$ is a function of the thermodynamic volume, $V$ [see Eq. (\ref%
{volrp})]. Following \cite{PVpapers}, we identify the geometric quantities $%
P $\ and $T$\ with physical pressure and temperature of system by using
dimensional analysis and $l_{P}^{n-1}=G_{n+1}\hbar /c^{3}$ as
\begin{equation}
\lbrack \mbox{Press}]=\frac{\hbar c}{l_{p}^{n-1}}[P],\quad \lbrack %
\mbox{Temp}]=\frac{\hbar c}{k}[T].  \label{dimless1}
\end{equation}%
Therefore, the physical pressure and physical temperature are given by
\begin{eqnarray}
\mbox{Press} &=&\frac{\hbar c}{l_{p}^{n-1}}P=\frac{\hbar c}{l_{p}^{n-1}}%
\frac{(n-1)T}{4r_{+}}+\dots  \notag \\
&=&\frac{k\mbox{Temp}(n-1)}{4l_{p}^{n-1}r_{+}}+\dots \;.  \label{dimless2}
\end{eqnarray}%
Now, one could compare them with the Van der Waals equation \cite{PVpapers},
and identify the specific volume $v$\ of the fluid with the horizon radius
as $v=\frac{4r_{+}l_{P}^{n-1}}{n-1}$, and in geometric units ($l_{P}=1$, $%
r_{+}=\left( n-1\right) v/4$), the equation of state (\ref{state}) can be
written in the following form
\begin{eqnarray}
P &=&\frac{T}{v}-\frac{(n-2)}{\pi (n-1)v^{2}}+\frac{1}{16\pi }\frac{\kappa
q^{2s}}{v^{2s(n-1)/(2s-1)}},  \label{StateV} \\
\kappa &=&\frac{4^{2s(n-1)/(2s-1)}(2s-1)\left( \frac{(n-1)(2s-n)^{2}}{%
(n-2)(2s-1)^{2}}\right) ^{s}}{(n-1)^{2s(n-1)/(2s-1)}},  \label{kappa}
\end{eqnarray}%
Considering Eq. (\ref{StateV}), we plot the $P-V$ isotherm diagram in Fig. %
\ref{FPV}. This figure shows that, similar to Van der Waals gas,
there is a critical point which is a point of inflection on the
critical isotherm. The pressure and volume at the critical point
are known as the critical pressure and the critical volume,
respectively. Above the critical point and for large volumes and
low pressures, the isotherms lose their inflection points and
approach equilateral hyperbolas, the so-called the isotherms of an
ideal gas. It is shown that the slope of the isotherm $P-V$
diagram passing through the critical point is zero. Furthermore,
as we mentioned before, the critical point is a point of
inflection on the critical isotherm, hence
\begin{eqnarray}
\frac{\partial P}{\partial v} &=&0,  \label{dpdv} \\
\quad \frac{\partial ^{2}P}{\partial v^{2}} &=&0.  \label{d2pdv2}
\end{eqnarray}%
Using Eqs. (\ref{dpdv}) and (\ref{d2pdv2}) with the equation of state (\ref%
{StateV}), we will be able to calculate the critical parameters
\begin{eqnarray}
v_{c} &=&\left[ \frac{\kappa s(n-1)^{2}(2ns-4s+1)q^{2s}}{16(n-2)(2s-1)^{2}}%
\right] ^{(2s-1)/[2(ns-3s+1)]},  \label{Vc} \\
T_{c} &=&\frac{4(n-2)(ns-3s+1)\left[ \frac{\kappa s(n-1)^{2}(2ns-4s+1)q^{2s}%
}{16(n-2)(2s-1)^{2}}\right] ^{(1-2s)/[2(ns-3s+1)]}}{\pi (n-1)(2ns-4s+1)},
\label{Tc} \\
P_{c} &=&\frac{(n-2)(ns-3s+1)}{\pi s(n-1)^{2}\left[ \frac{\kappa
s(n-1)^{2}(2ns-4s+1)q^{2s}}{16(n-2)(2s-1)^{2}}\right] ^{(2s-1)/(ns-3s+1)}}.
\label{Pc}
\end{eqnarray}%
These relations lead us to obtain the following universal ratio
\begin{equation}
{\rho }_{c}=\frac{P_{c}v_{c}}{T_{c}}=\frac{2ns-4s+1}{4s(n-1)}.
\label{UniversalRatio}
\end{equation}%
Note that for $s=-2/(n-5)$ with arbitrary spacetime dimensions, one can
recover the ratio $\rho _{c}=3/8$, characteristic for a Van der Waals gas.

\subsection{ Free energy}

Thermodynamic behavior of a system may be governed by the
thermodynamic potentials such as the free energy. It is known that
the free energy of a gravitational system may be obtained by
evaluating the Euclidean on-shell action. In order to calculate
it, we use the counterterm method for cancelling of divergences.
Furthermore, to make an action well-defined, one should add the
Gibbons-Hawking boundary term to the bulk action. In addition, in
order to fix charge on the boundary (working in canonical
ensemble) we should consider a boundary term for electromagnetic
field. So the total action is \cite{DEhShakVah}
\begin{equation}
I=I_{b}+I_{ct}-\frac{1}{8\pi }\int_{\partial M}d^{n}x~\sqrt{\gamma }~K-\frac{%
s}{4\pi }\int_{\partial M}d^{n}x~\sqrt{\gamma }(-\mathcal{F})^{s-1}~n_{\mu
}F^{\mu \nu }A_{\nu },  \label{FullAction}
\end{equation}%
where $I_{ct}$ is the counterterm action, and $\gamma _{ij}$ and $K$ denote
the induced metric and extrinsic curvature of the boundary. Using Eq. (\ref%
{FullAction}), it is straightforward to calculate the on-shell value of the
total action
\begin{equation}
I=\frac{\beta \omega _{n-1}}{16\pi }\left( 1-{\frac{r_{+}^{2}}{{l}^{2}}}+{%
\frac{\left( 2s-1\right) (2sn-4s+1)\Psi ^{s}r_{+}^{2}}{\left( n-1\right)
\left( n-2s\right) }}\right) r_{+}^{n-2},  \label{Onshell}
\end{equation}%
where
\begin{equation*}
\Psi =\left( \frac{n-1}{n-2}\right) \left( \frac{2s-n}{2s-1}\right)
^{2}q^{2}r_{+}^{-\frac{2(n-1)}{2s-1}}.
\end{equation*}%
and $\beta $ is the periodic Euclidean time which is related to
the inverse of Hawking temperature. Using the fact that $G=I\beta
^{-1}$ with Eq. (\ref{PLambda}), the (fixed charge) free energy in
the extended phase space may be written as
\begin{equation}
G(T,P)=\frac{\omega _{n-1}}{16\pi }\left( {1}-\frac{16\pi Pr_{+}^{2}}{n(n-1)}%
+{\frac{\left( 2s-1\right) (2sn-4s+1)\Psi ^{s}r_{+}^{2}}{\left( n-1\right)
\left( n-2s\right) }}\right) {r}_{+}^{n-2}.
\end{equation}

\begin{figure}[tbp]
$%
\begin{array}{cc}
\epsfxsize=7cm \epsffile{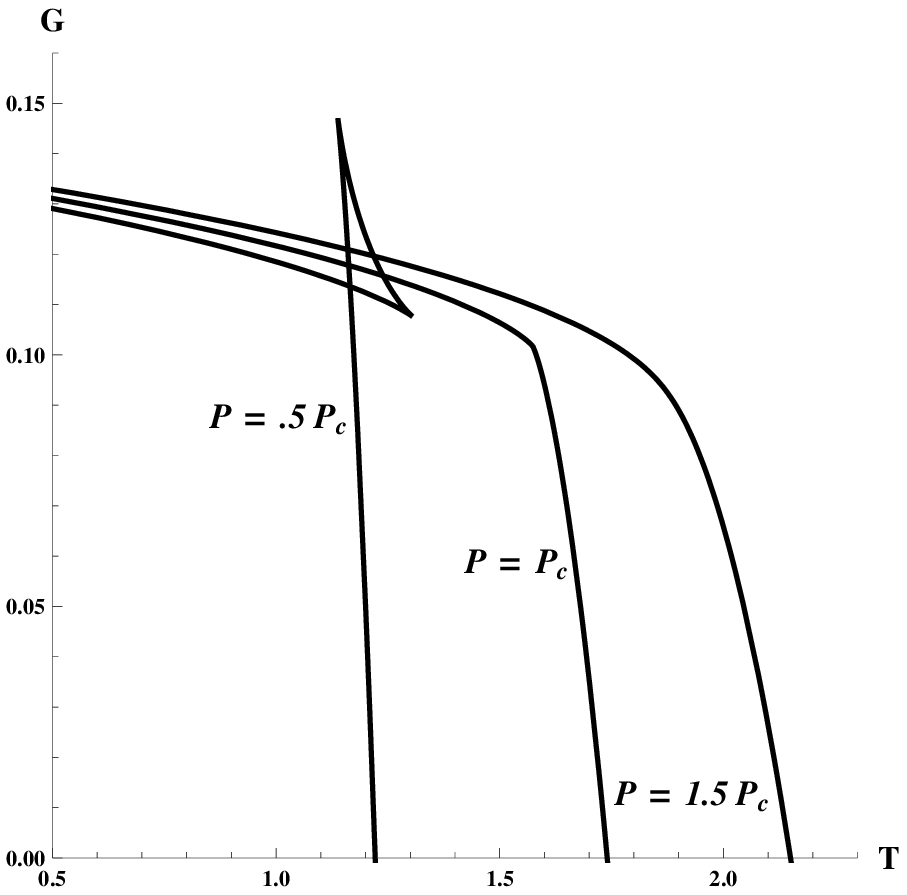} & \epsfxsize=7cm \epsffile{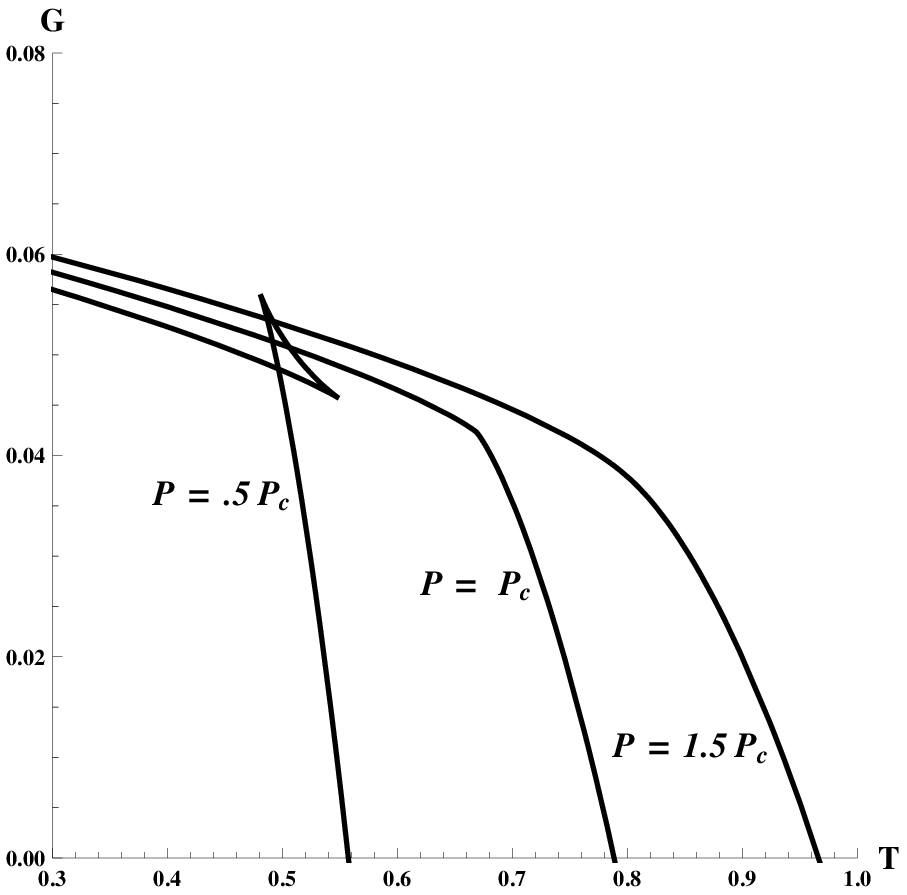}%
\end{array}
$%
\caption{ Free energy density ($\frac{G(T,P)}{\protect\omega _{n-1}}$) of
charged black holes with PMI source with respect to temperature for $q=1$, $%
s=\frac{6}{5}$ with $n=4$ (left) and $q=1$, $s=\frac{3}{4}$ with $n=3$
(right) . The characteristic swallowtail behavior is the signature of the
first order phase transition between large-small charged black holes (
analogous to the Van der Waals gas).}
\label{FG}
\end{figure}

\begin{figure}[tbp]
$%
\begin{array}{cc}
\epsfxsize=7cm \epsffile{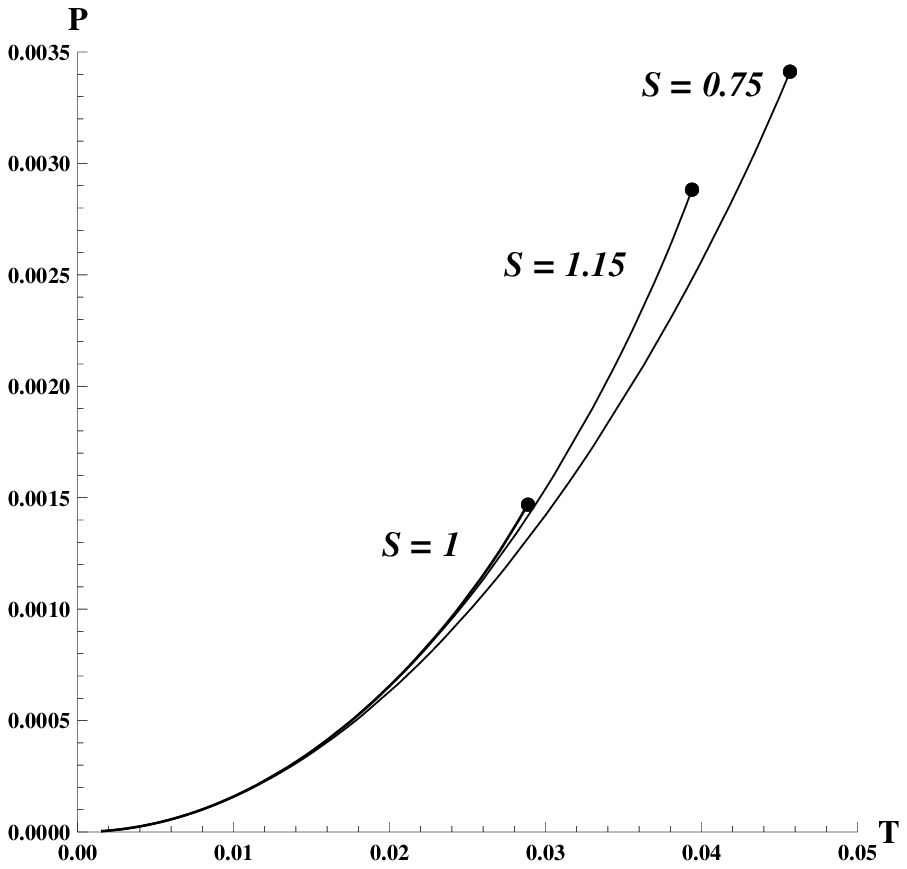} & \epsfxsize=7cm %
\epsffile{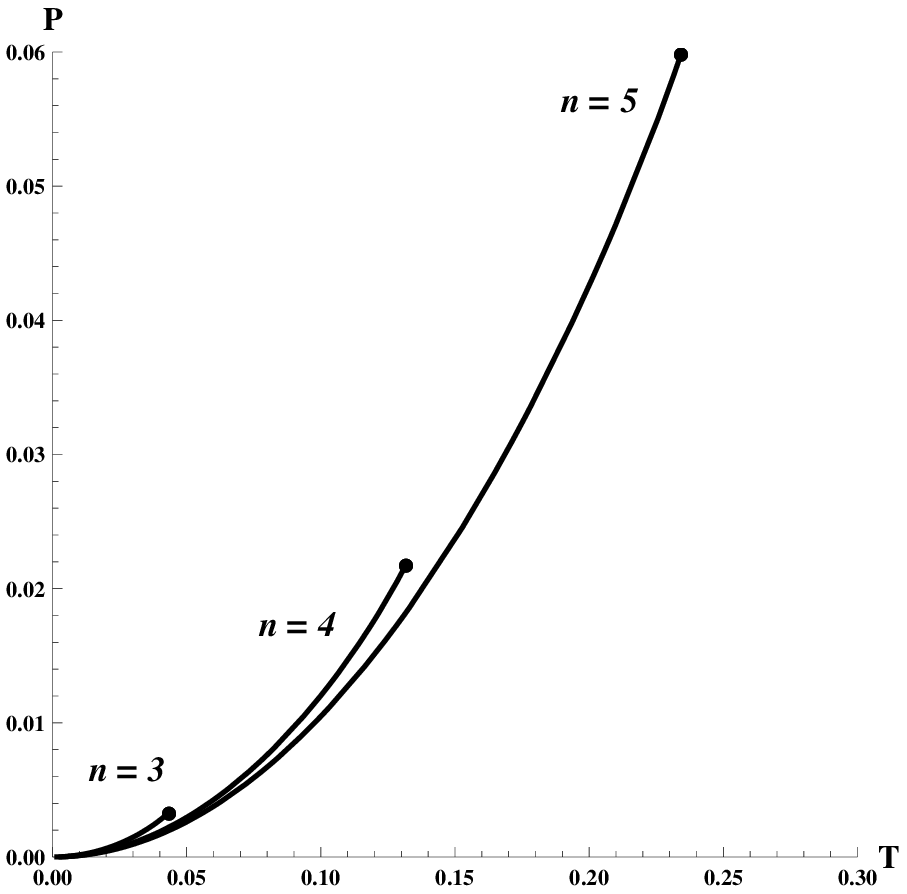}%
\end{array}
$%
\caption{Coexistence curves of large and small charged black holes
with PMI source for $q=1.5$ with $n=3$ (left) and $q=1$ with
$s=\frac{n+1}{4}$(right) . Critical points are denoted by the
small circle at the end of the coexistence curve. Above these
points, the phase transition does not occur. } \label{PT}
\end{figure}

The behavior of the free energy is displayed in Fig. \ref{FG}. In
this figure the characteristic swallowtail behavior of the free
energy shows the first order phase transition happen between large
and small charged black holes. Using the fact that the free
energy, temperature and the pressure of the system are constant
during the phase transition, one can plot the coexistence curve of
two phases large and small charged black holes in the PMI theory
(see Fig. \ref{PT}). Along this curve, small and large black holes
have alike temperature (horizon radii) and pressure.

\subsubsection{Critical exponents}

One of the most important characteristics of the phase transition
is the value of its critical exponents. So, following the approach
of \cite{PVnonlinear}, we calculate the critical exponents $\alpha
$, $\beta $, $\gamma $, $\delta $ for the phase transition of
$(n+1)$-dimensional charged black holes with an arbitrary $s$. In
order to obtain the critical exponent $\alpha $, we consider the
entropy of horizon $S$ and rewrite it in terms of $T$ and $V$. So
we have
\begin{equation}
S=S(T,V)=\Bigl[\omega _{n-1}\Bigl(nV\Bigr)^{n-1}\Bigr]^{\frac{1}{n}}.
\label{Ent}
\end{equation}%
Obviously, this is independent of $T$ and then the specific heat vanishes, ($%
C_{V}=0$), and hence $\alpha =0$. To obtain other exponents, we study
equation of state (\ref{StateV}) in terms of reduced thermodynamic variables
\begin{equation}
p=\frac{P}{P_{c}},\quad \nu =\frac{v}{v_{c}},\quad \tau =\frac{T}{T_{c}}.
\label{Reduced}
\end{equation}%
So, Eq. (\ref{StateV}) translates into the following reduced equation of
state
\begin{equation}
p=\frac{4(n-1)s\tau }{(2ns-4s+1)\nu }-\frac{n-1}{(ns-3s+1)\nu ^{2}}+\frac{%
(2s-1)^{2}}{(2ns-4s+1)(ns-3s+1)\nu ^{\frac{2s(n-1)}{2s-1}}}.  \label{statesd}
\end{equation}%
To study the recent equation, we will slightly generalize the
argument of \cite{PVnonlinear} for nonlinear Maxwell theory.
Indeed, we can rewrite the equation of state (\ref{statesd}) as
\begin{equation}
p=\frac{1}{\rho _{c}}\frac{\tau }{\nu }+f(\nu ,s),  \label{general}
\end{equation}%
where $\rho _{c}$\ stands for the critical ratio and
\begin{equation*}
f(\nu ,s)=\frac{1}{s\left( 1-4{\rho }_{c}\right) }\left( \frac{1}{\nu ^{2}}-%
\frac{\left( \frac{2s-1}{n-1}\right) ^{2}}{4s{\rho }_{c}\nu ^{\frac{2s(n-1)}{%
2s-1}}}\right) .
\end{equation*}%
The function $f(v,s)$ depends on $v$ and $s$ compared to \cite{PVnonlinear}
where it is independent of $s$. But as we will see the nonlinearity
parameter $s$ does not play any dramatic role and does not change critical
exponents. Following the method of Ref. \cite{PVnonlinear}, one may define
two new parameters $t$ and $\omega $
\begin{equation}
\tau =t+1,\quad \nu =(\omega +1)^{1/\epsilon },  \label{omegat}
\end{equation}%
where $\epsilon $ is a positive parameter. Now we can expand (\ref{statesd})
near the critical point to obtain
\begin{equation}
p=1+At-Bt\omega -C\omega ^{3}+O(t\omega ^{2},\omega ^{4}),
\label{generalexpansion}
\end{equation}%
with
\begin{equation}
A=\frac{1}{\rho _{c}},\quad B=\frac{1}{\epsilon \rho _{c}},\quad C=\frac{%
2s(n-1)}{3\epsilon ^{3}(2s-1)}.  \label{ABC}
\end{equation}%
We consider a fixed $t<0$ and differentiate the Eq. (\ref{generalexpansion})
to obtain
\begin{equation}
dP=-P_{c}(Bt+3 C \omega ^{2})d\omega .  \label{dPgeneral}
\end{equation}%
Now, we denote the volume of small and large black holes with $\omega _{s}$
and $\omega _{l}$, respectively, and apply the Maxwell's equal area law. One
obtains
\begin{eqnarray}
p &=&1+At-Bt\omega _{l}-C\omega _{l}^{3}=1+At-Bt\omega _{s}-C\omega _{s}^{3}
\notag \\
0 &=&\int_{\omega _{l}}^{\omega _{s}}\omega dP.
\end{eqnarray}
This equation leads to a unique non-trivial solution
\begin{equation}
\omega _{s}=-\omega _{l}=\sqrt{\frac{-Bt}{C}},
\end{equation}
and therefore we can find
\begin{equation}
\eta =V_{c}(\omega _{l}-\omega _{s})=2V_{c}\omega _{l}\propto \sqrt{-t}\quad
\Rightarrow \quad \beta =\frac{1}{2}.
\end{equation}
Now, we should calculate the next exponent, $\gamma $. In order to obtain
it, one should consider Eq. (\ref{generalexpansion}). After some
manipulation one can obtain
\begin{equation}
\kappa _{T}=-\frac{1}{V}\frac{\partial V}{\partial P}\Big |_{T}\propto \frac{%
1}{P_{c}}\frac{1}{Bt}\quad \Rightarrow \quad \gamma =1.
\end{equation}%
Next, we calculate the final exponent, $\delta $. To do this, we should
obtain the shape of the critical isotherm $t=0$ (\ref{generalexpansion}),
i.e.,
\begin{equation}
p-1=-C\omega ^{3}\quad \Rightarrow \quad \delta =3.
\end{equation}%
We conclude that the thermodynamic exponents associated with the nonlinear
charged black holes in any dimension $n\geq 3$ with arbitrary nonlinearity
parameter, $s\neq n/2$, coincide with those of the Van der Waals fluid (the
same as critical exponents of the linear Maxwell case).

\subsection{Equation of state for the BTZ-like black holes}

So far, we have investigated the phase transition of black holes
in the presence of nonlinear PMI source with the nonlinearity
$s\neq n/2$. Interestingly, for $s=n/2$, the solutions (the
so-called BTZ-like black holes) have different properties. In
other words, the solutions for $s=n/2$ are not the special limit
of the solutions for general $s$. In fact, the solutions of
$s=n/2$ are completely special and differ from the solutions of
other values of $s$. As we will see, for $s=n/2$ the charge term
in metric
function is logarithmic and the electromagnetic field is proportional to $%
r^{-1}$ (logarithmic gauge potential). In other words, in spite of some
differences, this special higher dimensional solution has some similarity
with the charged BTZ solution and reduces to the original BTZ black hole for $%
n=2 $.

Considering the metric (\ref{Metric}) and the field equations of the bulk
action (\ref{Action}) with $s=n/2$, we can find that the metric function $%
f(r)$ and the gauge potential may be written as
\begin{eqnarray}
f(r) &=&1+\frac{r^{2}}{l^{2}}-\frac{m}{r^{n-2}}-\frac{2^{n/2}q^{n}}{r^{n-2}}%
\ln \left( \frac{r}{l}\right) ,  \label{Vbtz} \\
A &=&q\ln \left( \frac{r}{l}\right) dt,  \label{Abtz}
\end{eqnarray}
Straightforward calculations show that BTZ-like spacetime has a
curvature singularity located at $r=0$, which is covered with an
event horizon. The temperature of this black hole can be obtained
as \cite{BTZlike}
\begin{equation}
T=\frac{n-2}{4\pi r_{+}}\left( 1+\frac{n}{n-2}\frac{r_{+}^{2}}{l^{2}}-\frac{%
2^{n/2}q^{n}}{(n-2)r_{+}^{n-2}}\right) ,  \label{TBTZ}
\end{equation}

In this section, we will investigate the analogy of the liquid--gas phase
transition of the Van der Waals fluid with the\ phase transition in BTZ-like
black hole solutions \cite{BTZlike}. Following the same approach and using
Eqs. (\ref{PLambda}) and (\ref{TBTZ}) for a fixed charge $Q$, we obtain
\begin{equation}
P=\frac{(n-1)}{4r_{+}}T-\frac{(n-1)(n-2)}{16\pi r_{+}^{2}}+\frac{1}{16\pi }%
\frac{2^{n/2}(n-1)q^{n}}{r_{+}^{n}}.  \label{P1}
\end{equation}

Using Eqs. (\ref{dimless1}) and (\ref{dimless2}) with the fact that in
geometric units $v=\frac{4r_{+}}{n-1}$, Eq. (\ref{P1}) may be rewritten as
\begin{eqnarray}
P &=&\frac{T}{v}-\frac{(n-2)}{\pi (n-1)v^{2}}+\frac{1}{16\pi }\frac{\kappa
^{\prime }q^{2s}}{v^{n}},  \label{P2} \\
\kappa ^{\prime } &=&\frac{2^{5n/2}}{(n-1)^{n-1}}.
\end{eqnarray}

\begin{figure}[tbp]
$%
\begin{array}{cc}
\epsfxsize=7cm \epsffile{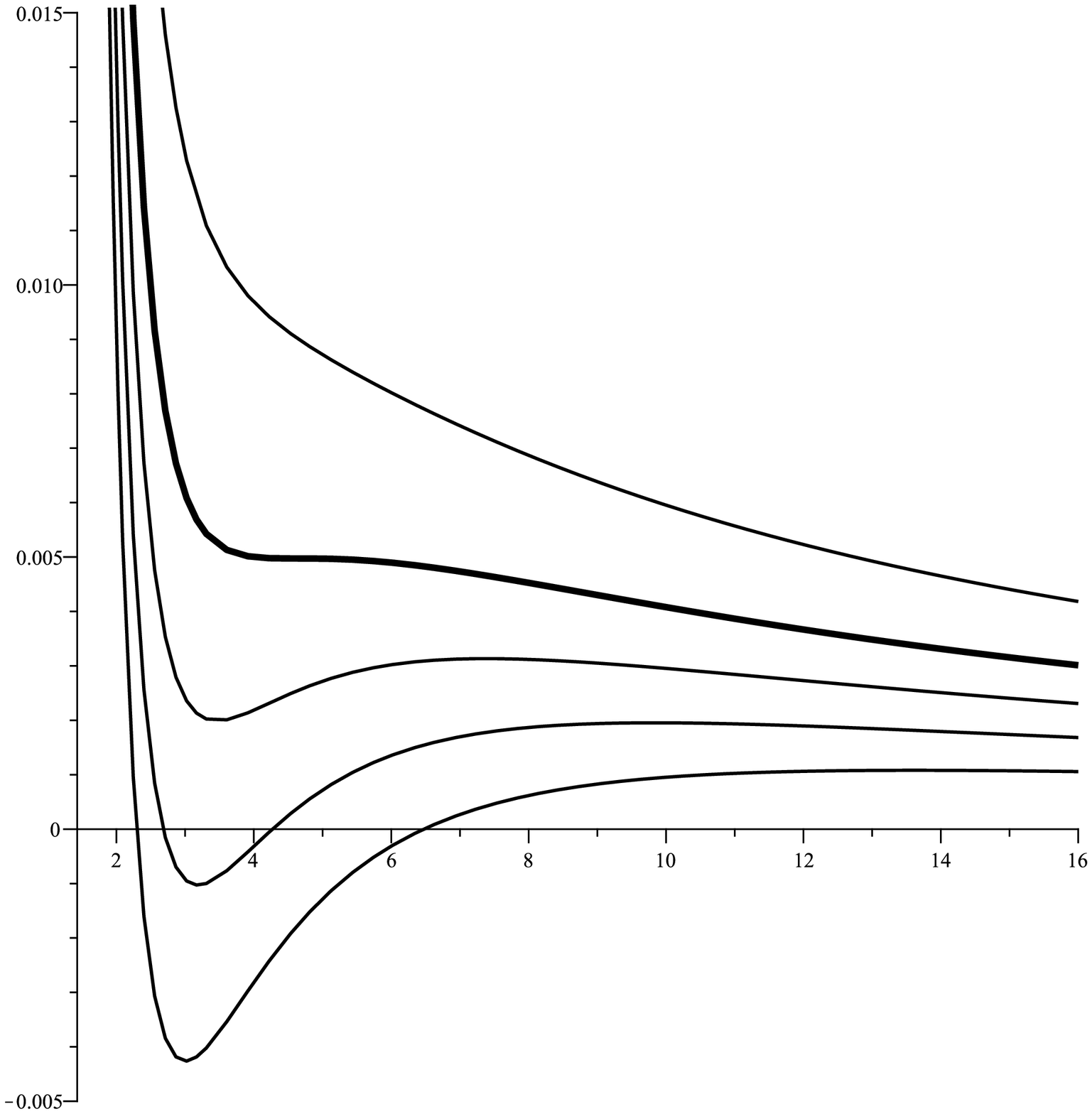} & \epsfxsize=7cm \epsffile{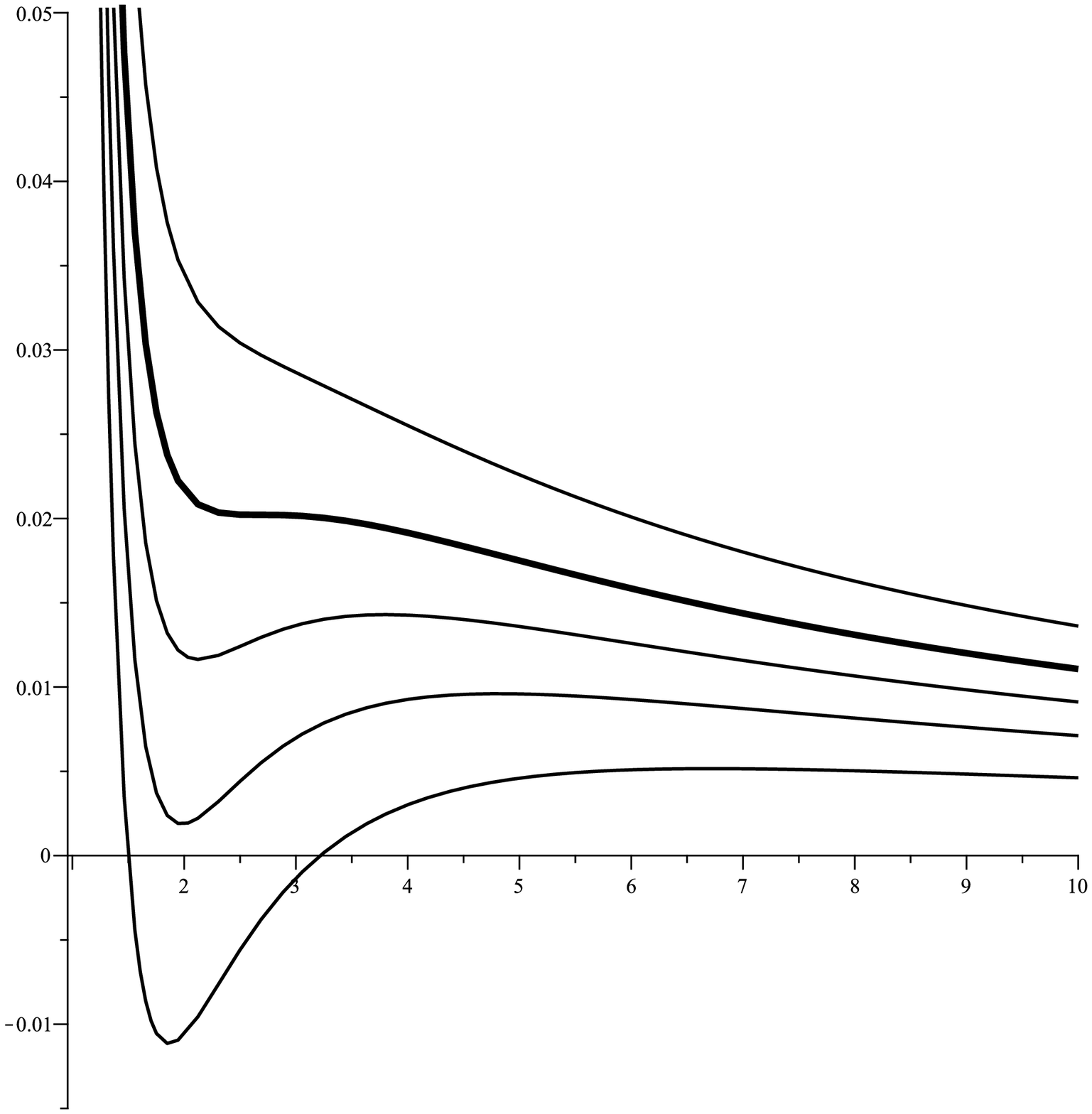}%
\end{array}
$%
\caption{$P-V$ diagram of BTZ-like black holes for $n=4$ (left)
and $n=5$ (right). The temperature of isotherms decreases from top
to bottom. The bold line is the critical isotherm diagram.}
\label{FPVbtz}
\end{figure}

Now, we plot the isotherm $P-V$ diagram in Fig. \ref{FPVbtz}. The
behavior of these plots is the same as the Van der Waals gas. In
order to find the critical quantities, one may use Eqs.
(\ref{dpdv}) and (\ref{d2pdv2}). These relations help us to obtain
\begin{eqnarray}
v_{c} &=&\left[ \frac{\kappa ^{\prime }n(n-1)^{2}q^{n}}{32(n-2)}\right]
^{1/(n-2)},  \label{Vc2} \\
T_{c} &=&\frac{2(n-2)^{2}\left[ \frac{\kappa ^{\prime }n(n-1)^{2}q^{n}}{%
32(n-2)}\right] ^{-1/(n-2)}}{\pi (n-1)^{2}},  \label{Tc2} \\
P_{c} &=&\frac{(n-2)^{2}}{\pi n(n-1)\left[ \frac{\kappa ^{\prime
}n(n-1)^{2}q^{n}}{32(n-2)}\right] ^{2/(n-2)}}.  \label{Pc2}
\end{eqnarray}%
Having the critical quantities at hand, we are in a position to obtain the
following universal ratio
\begin{equation}
{\rho }_{c}=\frac{P_{c}v_{c}}{T_{c}}=\frac{n-1}{2n}.  \label{ratio}
\end{equation}%
It is notable that only for $n=4$ ($5$-dimensional BTZ-like black holes),
one can recover the ratio $\rho _{c}=3/8$, characteristic for a Van der
Waals gas, where for higher dimensional Reissner--Nordstr\"{o}m black holes,
this ratio has been recovered only for $4$-dimensions \cite{PVnonlinear}. In
addition, considering $s=n/2$ in Eq. (\ref{UniversalRatio}), one can obtain $%
{\rho }_{c}$ of the BTZ-like black holes.

\subsubsection{Critical exponents of BTZ-like black holes}

In order to obtain the critical exponents of the phase transition,
we follow the same procedure of \cite{PVnonlinear}. The entropy of
horizon $S(T,V)$ is the same as Eq. (\ref{Ent}), which is
independent of $T$ and hence $\alpha =0 $. In addition we can use
Eq. (\ref{Reduced}) and rewrite Eq. (\ref{P2}) in the following
form
\begin{equation}
p=\frac{2n\tau }{(n-1)\nu }-\frac{n}{(n-2)\nu ^{2}}+\frac{2}{(n-1)(n-2)\nu
^{n}}.  \label{statesd2}
\end{equation}

It is straightforward to show that the thermodynamic exponents associated
with the BTZ-like black holes in arbitrary dimension, coincide with those of
the Van der Waals fluid (the same as critical exponents of the PMI case).

\section{Grand canonical ensemble}\label{GCano}

In addition to the canonical ensemble, one can work with a fixed
electric potential at infinity. The ensemble of this fixed
intensive quantity translates into the grand canonical ensemble.
It is worthwhile to note that, for linear Maxwell field, the
criticality cannot happen in the grand canonical ensemble
\cite{PVpapers}.

\subsection{Equation of state}

In this section, we study the critical behavior of charged black holes in
the grand canonical (fixed $\Phi $) ensemble. We take $q=\Phi
r_{+}^{(n-2s)/(2s-1)}$ with $v=\frac{4r_{+}}{n-1}$ to rewrite Eq. (\ref%
{state}) in the following form
\begin{equation}
P=\frac{T}{v}\mathbf{-}\frac{(n-2)}{(n-1)\pi v^{2}}\mathbf{+}\frac{2s-1}{%
16\pi }\left( \frac{4\sqrt{2}(n-2s)\Phi }{(2s-1)(n-1)v}\right) ^{2s}\mathbf{,%
}  \label{State Grand}
\end{equation}
In the Maxwell theory ($s=1$) Eq. (\ref{State Grand}) reduces to
\begin{equation}
Pv^{2}=Tv-\frac{(n-2)}{(n-1)\pi }+\frac{2}{\pi }\frac{(n-2)^{2}}{(n-1)^{2}}%
\Phi ^{2}\mathbf{.}
\end{equation}

\begin{figure}[tbp]
$%
\begin{array}{cc}
\epsfxsize=7cm \epsffile{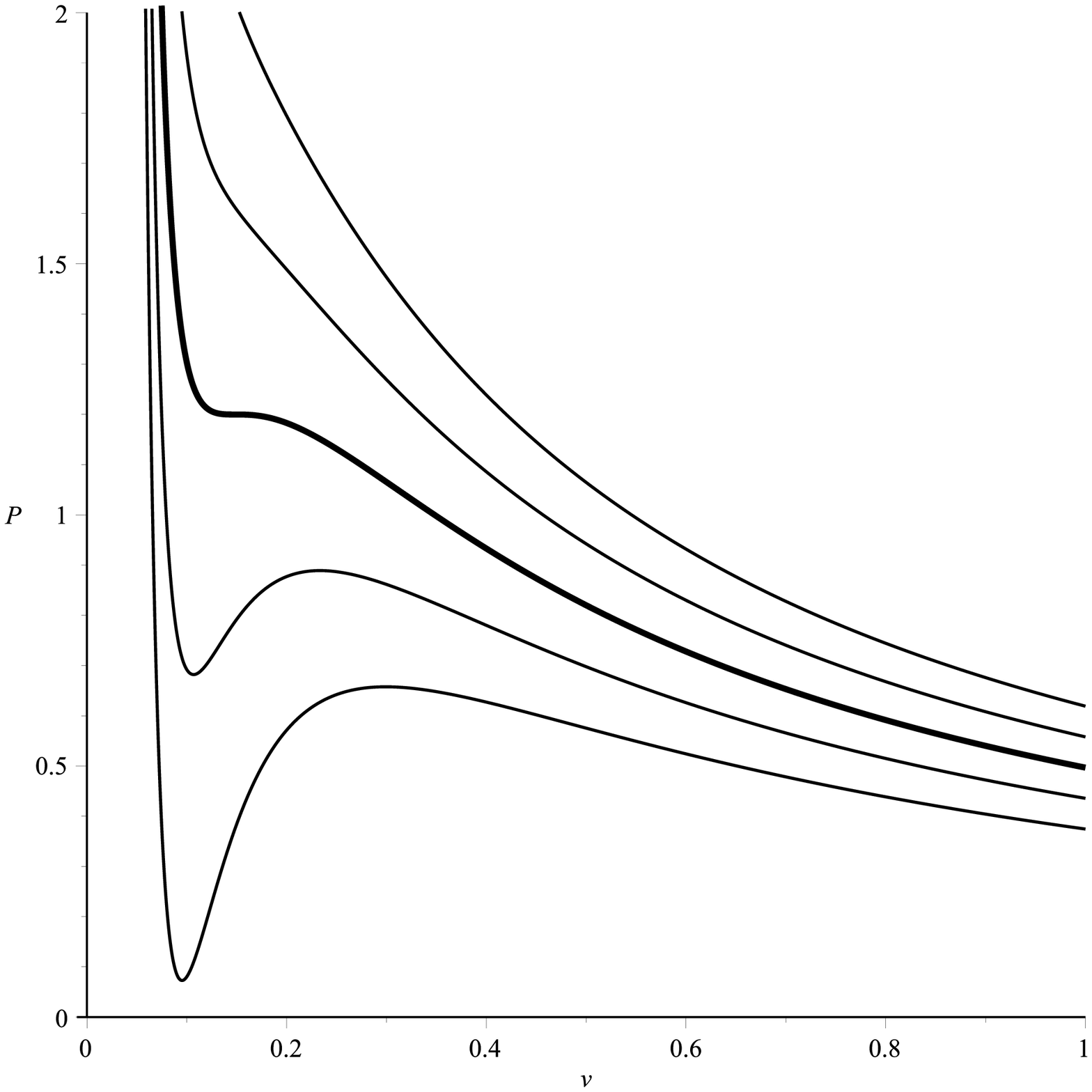} & \epsfxsize=7cm %
\epsffile{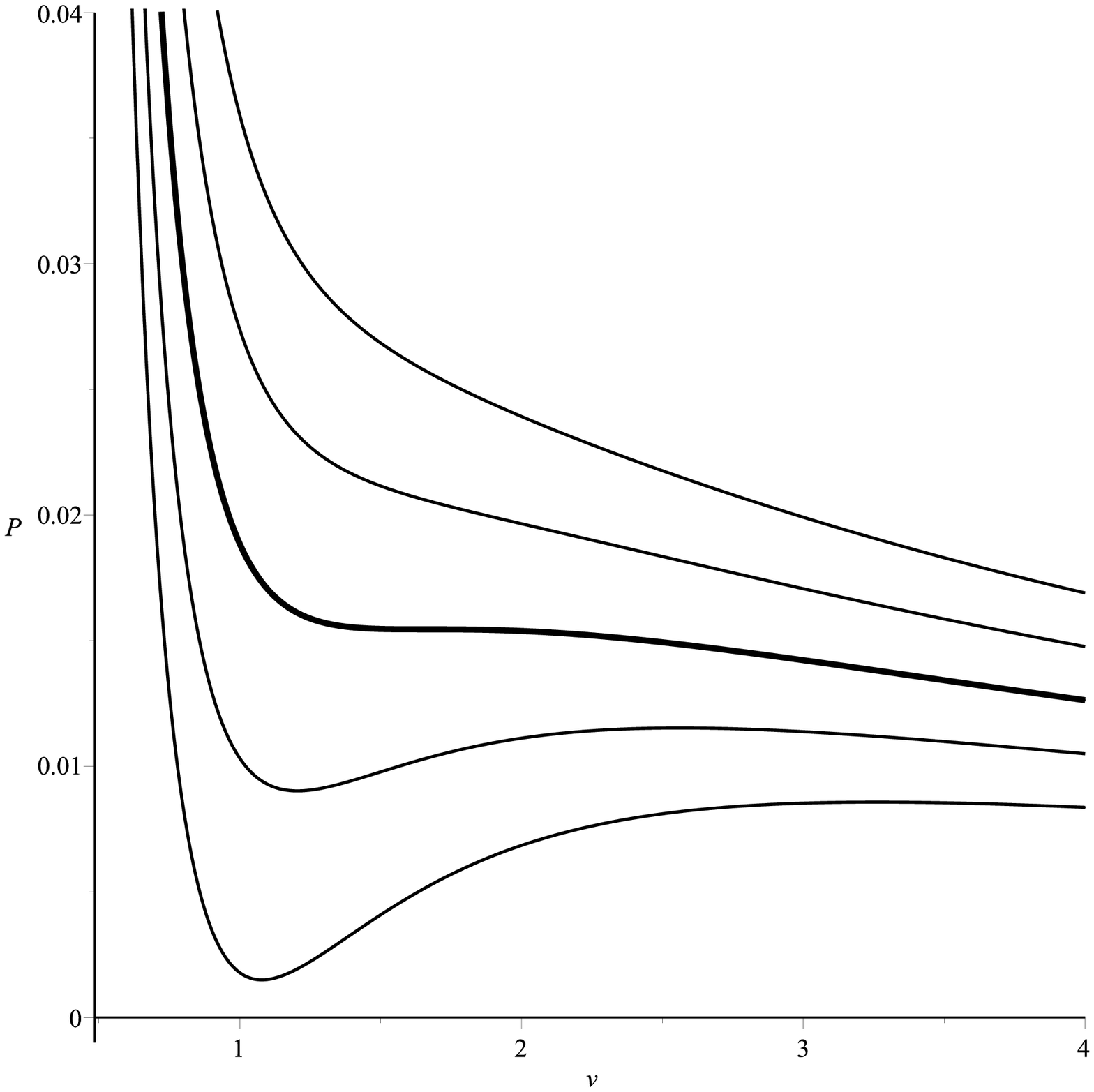}%
\end{array}
$%
\caption{$P-V$ diagram of charged AdS black holes in PMI for
$s=\frac{6}{5}$ with $n=3$ (left) and $s=\frac{5}{4}$ with $n=4$
(right). The temperature of isotherms decreases from top to
bottom. The bold line is the critical isotherm diagram.}
\label{PotPV}
\end{figure}

Clearly, this is a quadratic equation and does not show any
criticality. Interestingly, in contrast to the Maxwell field
($s=1$), the PMI theory admits phase transition in this ensemble
and one can study the fixed potential $P-V$ phase transition of
the black holes in extended phase space. However, the general
behavior of the isotherm $P-V$ diagram for fixed potential is same
as fixed charge ensemble as displayed in Fig \ref{PotPV}. Applying
Eqs. (\ref{dpdv}) and (\ref{d2pdv2}) to the equation of state, it
is easy to calculate the critical point in the grand canonical
ensemble
\begin{eqnarray}
v_{c} &=&\frac{4\sqrt{2}(n-2s)}{(2s-1)(n-1)}\left[ \frac{2s(2s-n)^{2}}{%
(n-2)(n-1)}\right] ^{\frac{1}{2(s-1)}}\Phi ^{\frac{s}{s-1}}, \\
T_{c} &=&\frac{(s-1)(n-2)}{\pi (n-2s)}\left[ \frac{(n-2)(n-1)}{%
s2^{s}(n-2s)^{2}}\right] ^{\frac{1}{2(s-1)}}\Phi ^{\frac{-s}{s-1}}, \\
P_{c} &=&\frac{(s-1)(2s-1)^{2}2^{(4-5s)/(s-1)}}{s\pi }\left[ \frac{%
(n-2s)^{2}s^{\frac{1}{s}}}{(n-1)(n-2)}\right] ^{\frac{-s}{s-1}}\Phi ^{\frac{%
-2s}{s-1}}.
\end{eqnarray}%
One must consider that in contrast to the canonical ensemble for $s>\frac{n}{2}$%
\ or $s<1~$the $v_{c}$,~$T_{c}$\ or $P_{c}~$\ take the negative value so
there is not any physical phase transition in these cases. Using the values
of $v_{c},T_{c\text{ }}$and $P_{c}$\ we will be able to obtain the following
universal ratio%
\begin{equation}
{\rho }_{c}=\frac{P_{c}v_{c}}{T_{c}}=\frac{2s-1}{4s}.  \label{ratio2}
\end{equation}%
Note that here this universal ratio is independent of $n$ and\ for $s=2$ one
can recover the ratio $\rho _{c}=3/8$, characteristic for a Van der Waals
gas. Although Eq. (\ref{ratio2}) does not depend on the spacetime
dimensions, but one can take $s=n/2$ to recover Eq. (\ref{ratio}) of
BTZ-like black holes. Furthermore, it is interesting to mention that one can
set $n=2s$ in Eq. (\ref{UniversalRatio}) to obtain Eq. (\ref{ratio2}).

\subsection{ Free energy}

\begin{figure}[tbp]
$%
\begin{array}{cc}
\epsfxsize=7cm \epsffile{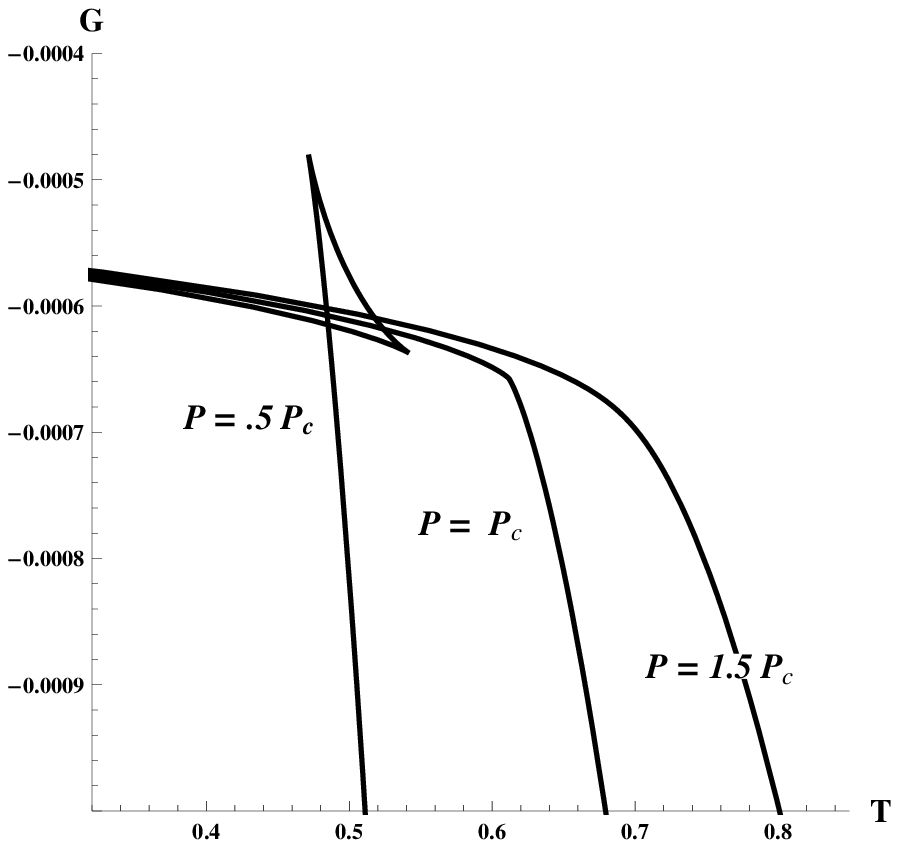} & \epsfxsize=7cm %
\epsffile{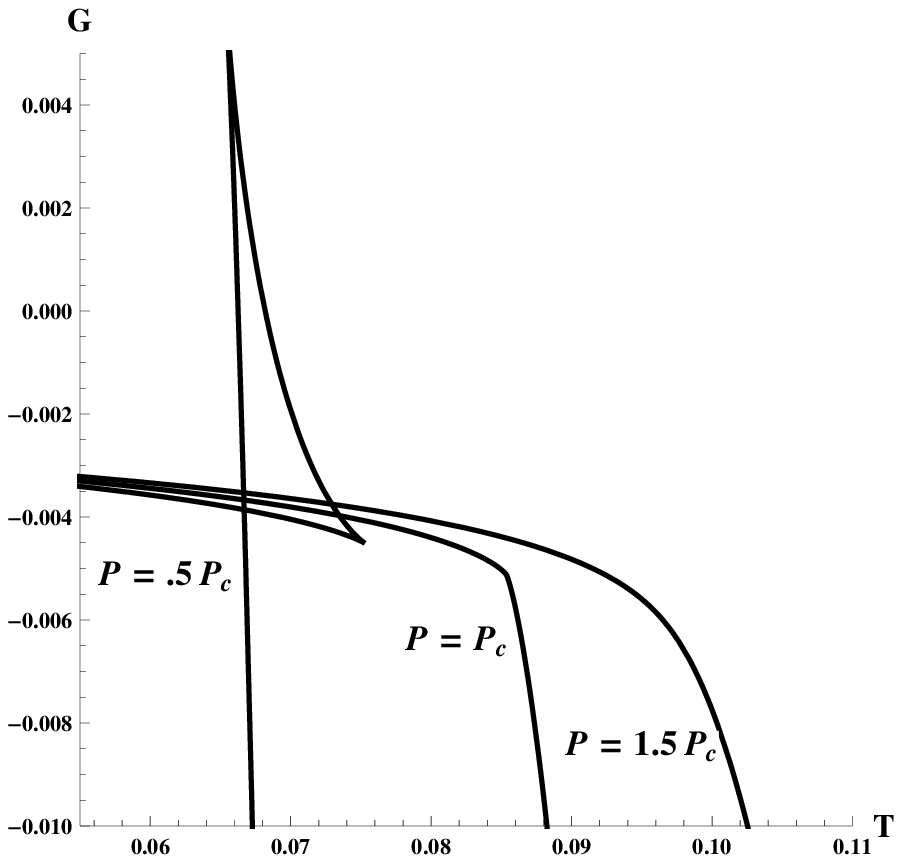}%
\end{array}
$%
\caption{ Free energy density ($\frac{G(T,P)}{\protect\omega _{n-1}}$) of
charged black holes with PMI source with respect to temperature for $\Phi =1$%
, $s=\frac{6}{5}$ with $n=3$ (left) and $\Phi =1$, $s=\frac{5}{4}$ with $n=4$
(right).}
\label{PotG}
\end{figure}

By ignoring the surface term of PMI and fixing the potential on the boundary $%
\delta A_{\mu }|_{\partial M}=0$, one can find the on-shell action
correspondence to free energy in the grand canonical ensemble. So,
we take the action as follows
\begin{equation}
I=I_{b}+I_{ct}-\frac{1}{8\pi }\int_{\partial M}d^{n}x~\sqrt{\gamma }~K.
\end{equation}
Now we can find the free energy $G=I\beta ^{-1}$ as
\begin{equation}
G(T,P)=\frac{\omega _{n-1}}{16\pi }\left( {1}-\frac{16\pi Pr_{+}^{2}}{n(n-1)}%
+{\frac{2^{s}\left( 2s-1\right) ^{2-2s}\Phi ^{2s}r_{+}^{2-2s}}{\left(
n-1\right) \left( 2s-n\right) ^{1-2s}}}\right) {r}_{+}^{n-2}.
\end{equation}
Now, we are looking for the phase transition. We plot figure
\ref{PotG} and find that, in contrast to Maxwell case, there is a
first order phase transition. In other words, the nonlinearity
parameter, $s$, affects the existence of the phase transition in
the grand canonical ensemble.

\section{Concluding Remarks}

In this paper, we have considered the cosmological constant and its
conjugate quantity as thermodynamic variables and investigated the
thermodynamic properties of a class of charged black hole solutions. At the
first step, we have introduced the black hole solutions of the Einstein-$%
\Lambda $ gravity in the presence of the PMI source.

Then, we have used the Hawking temperature as an equation of state
and calculated the the critical parameters, $T_{c}$, $v_{c}$ and
$P_{c}$. We have plotted the isotherm diagram ($P$--$V$) of
charged black holes in PMI theory and found that the total
behavior of this diagram is the same as that of the Van der Waals
gas. Also, we have obtained the free energy of a gravitational
system through the use of Euclidean on-shell action to investigate
its thermodynamics behavior.

Furthermore, we have calculated the critical exponents of the phase
transition and concluded that the thermodynamic exponents associated with
the nonlinear charged black holes in arbitrary dimension coincide with those
of the Van der Waals fluid (the mean field theory).

Also, we have applied the same procedure for the BTZ-like black
holes to obtain their phase transition. Calculations showed that
thermodynamic behaviors of BTZ-like black holes are the same as
PMI ones.

Moreover, we have studied the grand canonical ensemble in which
the potential, instead of charge, should be fixed on the boundary.
In contrast to the Maxwell case \cite{PVpapers}, here one sees a
phase transition. We have also computed the universal ratio
$\frac{P_{c}v_{c}}{T_{c}}=\frac{2s-1}{4s}$ and found that it does
not depend on the spacetime dimensions.

Finally, we have found that, $v_{c}$, $T_{c}$ and $P_{c}$ have
different dependencies of $n$ and $s$ for PMI and BTZ-like in the
canonical ensemble, but when $n=2s$ both of them reduce to the
universal ratio
$\rho_{c}=\frac{P_{c}v_{c}}{T_{c}}=\frac{2s-1}{4s}$ which we have
found in the grand canonical ensemble.

It is interesting to investigate underlining reasons for the
mentioned universality and figure out why this ratio in the grand
canonical ensembles is independent of spacetime dimensions.
Moreover, in the statistical physics, it is known that a
universality class of criticality is characterized by dimensions
of space, order parameters and fluctuations \cite{Gold}. However,
it is not clear which features of gravitational theories and black
holes determine the universality class of phase transition or
modify critical exponents. Clearly as we found in this paper and
have been shown in \cite{PVnonlinear}, the critical exponents do
not change by crucial modifications of matter fields (such as PMI
modification) or changing the space-time dimensions. In addition,
it seems the geometry of spacetime is also irrelevant to critical
exponents as one can see in the slowly rotating black holes
\cite{PVnonlinear}. Interestingly, these critical exponents remain
unchanged in the mean field class even when one considers some
corrections to gravity action \cite{GB}. Besides, it is worthwhile
to think about whether there is any holographic interpretation for
the extended phase space thermodynamics and the universal
classification. Perhaps a holographic approach helps us to have a
better understanding of this problem. We leave the study of these
interesting questions for future studies.

\section{Acknowledgement}

We thank an anonymous referee for useful comments. M. H. Vahidinia
would like to thank A. Montakhab, S. Jalali, P. Manshour and A.
Moosavi for useful discussions. S. H. Hendi wishes to thank Shiraz
University Research Council. This work has been supported
financially by Research Institute for Astronomy \& Astrophysics of
Maragha (RIAAM), Iran.

%%%%%%%%%%%%%%%%%%%%%%%%%%%%%%%%%%%%%%%%%%%%%%%%%%%%%%%%%%%%%%%%%%%%%%%%%%%%%%%%%%%%%%%%%%%%%%%%%%%%%%%%%%%%%%%%%%%%%%%%%%%%%%%%%%%%%%%%%
%%%%%%%%%%%%%%%%%%%%%%%%%%%%%%%%%%%%%%%%%%%%%%%%%%%%%%%%%%%%%%%%%%%%%%%%%%%%%%%%%%%%%%%%%%%%%%%%%%%%%%%%%%%%%%%%%%%%%%%%%%%%%%%%%%%%%%%%%%


\begin{thebibliography}{99}
\bibitem{Gibbons1} G. Gibbons, R. Kallosh and Barak Kol, Phys. Rev. Lett.
\textbf{77}, 4992 (1996).

\bibitem{BrownPLB1987} J.D. Brown and C. Teitelboim, Phys. Lett. B \textbf{%
195}, 177 (1987).

\bibitem{CaldarelliCQG2000} M. M. Caldarelli, G. Cognola and D. Klemm,
Class. Quantum Gravit. \textbf{17}, 399 (2000).

\bibitem{RayCQG2009} D. Kastor, S. Ray and J. Traschen, Class. Quantum
Gravit. \textbf{26}, 195011 (2009).

\bibitem{Gibbons2} M. Cvetic, G. Gibbons, D. Kubiznak and C. Pope, Phys.
Rev. D \textbf{84}, 024037 (2011).

\bibitem{PVpapers} B. P. Dolan, Class. Quantum Gravit. \textbf{28}, 125020
(2011);

B. P. Dolan, Class. Quantum Gravit. \textbf{28}, 235017 (2011);

D. Kubiznak and R. B. Mann, JHEP \textbf{07}, 033 (2012);

B. P. Dolan, [arXiv:1209.1272].

\bibitem{Wu2012} C. Niu, Y. Tian and X. Wu, Phys. Rev. D \textbf{85}, 024017
(2012).

\bibitem{MyersPRD1999} A. Chamblin, R. Emparan, C. Johnson and R. Myers,
Phys. Rev. D \textbf{60}, 064018 (1999);

A. Chamblin, R. Emparan, C. Johnson and R. Myers, Phys. Rev. D \textbf{60},
104026 (1999).

\bibitem{Banerjee:2011au} R. Banerjee and D. Roychowdhury, JHEP \textbf{1111}%
, 004 (2011).

\bibitem{Dirac} P. A. M. Dirac, \emph{Lectures on Quantum Mechanics},
Yeshiva University, Belfer Graduate School of Science, New York
(1964).

\bibitem{Bialynicka} Z. Bialynicka-Birula and I. Bialynicka-Birula, Phys.
Rev. D \textbf{2}, 2341 (1970).

\bibitem{Heisenberg} W. Heisenberg and H. Euler, Z. Phys. \textbf{98}, 714
(1936); \emph{Translation by}: W. Korolevski and H. Kleinert, \emph{%
Consequences of Dirac's Theory of the Positron}, [physics/0605038];

H. Yajima and T. Tamaki, Phys. Rev. D \textbf{63}, 064007 (2001).

\bibitem{Delphenich} D. H. Delphenich, \emph{Nonlinear electrodynamics and
QED}, [arXiv: hep-th/0309108];

D. H. Delphenich, \emph{Nonlinear optical analogies in quantum
electrodynamics}, [arXiv: hep-th/0610088].

\bibitem{Schwinger} J. Schwinger, Phys. Rev. \textbf{82}, 664 (1951).

\bibitem{Stehle} P. Stehle and P. G. DeBaryshe, Phys. Rev. \textbf{152},
1135 (1966).

\bibitem{ThermoBI} Y. S. Myung, Y. W. Kim and Y. J. Park, Phys. Rev. D
\textbf{78}, 044020 (2008);

S. Fernando, Phys. Rev. D \textbf{74}, 104032 (2006);

O. Miskovic and R. Olea, Phys. Rev. D \textbf{77}, 124048 (2008);

Y. S. Myung, Y. W. Kim and Y. J. Park, Phys. Rev. D \textbf{78}, 084002
(2008);

R. Banerjee and D. Roychowdhury, Phys. Rev. D \textbf{85}, 044040 (2012);

R. Banerjee and D. Roychowdhury, Phys. Rev. D \textbf{85}, 104043 (2012).

\bibitem{Sun08} R. G. Cai and Y. W. Sun, JHEP \textbf{09}, 115 (2008);

X. H. Ge, Y. Matsuo, F. W. Shu, S. J. Sin and T. Tsukioka, JHEP \textbf{10},
009 (2008).

\bibitem{AdSCFTBI} S. Gangopadhyay and D. Roychowdhury, JHEP \textbf{1205},
002 (2012);

D. Roychowdhury, Phys. Rev. D \textbf{86}, 106009 (2012);

S. Gangopadhyay and D. Roychowdhury, JHEP \textbf{1205}, 156 (2012).

\bibitem{PVnonlinear} S. Gunasekaran, R. B. Mann and D. Kubiznak,
[arXiv:1208.6251].

\bibitem{Kats} Y. Kats, L. Motl and M. Padi, JHEP \textbf{0712}, 068 (2007);

D. Anninos and G. Pastras, JHEP \texttt{0907}, 030 (2009);

R. G. Cai, Z. Y. Nie and Y. W. Sun, Phys. Rev. D \textbf{78}, 126007 (2008);

N. Seiberg and E. Witten, JHEP \textbf{09}, 032 (1999);

E. Fradkin and A. Tseytlin, Phys. Lett. B \textbf{163}, 123 (1985);

R. Matsaev, M. Rahmanov and A. Tseytlin, Phys. Lett. B \textbf{193}, 205
(1987);

E. Bergshoff, E. Sezgin, C. Pope and P. Townsend, Phys. Lett. B \textbf{188}%
, 70 (1987);

A. Tseytlin, Nucl. Phys. B \textbf{276}, 391 (1986);

D.J. Gross and J. H. Sloan, Nucl. Phys. B \textbf{291}, 41 (1987).

\bibitem{PMIpapers1} M. Hassaine and C. Martinez, Phys. Rev. D \textbf{75},
027502 (2007);

S. H. Hendi and H. R. Rastegar-Sedehi, Gen. Relativ. Gravit. \textbf{41},
1355 (2009);

S. H. Hendi, Phys. Lett. B \textbf{677}, 123 (2009);

\bibitem{PMIpapers2} M. Hassaine and C. Martinez, Class. Quantum Gravit.
\textbf{25}, 195023 (2008);

H. Maeda, M. Hassaine and C. Martinez, Phys. Rev. D \textbf{79}, 044012
(2009);

S. H. Hendi and B. Eslam Panah, Phys. Lett. B \textbf{684}, 77 (2010);

S. H. Hendi, Prog. Theor. Phys. \textbf{124}, 493 (2010);

S. H. Hendi, Eur. Phys. J. C \textbf{69}, 281 (2010);

S. H. Hendi, Phys. Rev. D \textbf{82}, 064040 (2010).

\bibitem{AdSCFTPMI} J. Jing, Q. Pan and S. Chen, JHEP \textbf{1111}, 045
(2011);

D. Roychowdhury, Phys. Lett. B \textbf{718}, 1089 (2013).

\bibitem{CIMFR} S. H. Hendi, Phys. Lett. B \textbf{690}, 220 (2010);

\bibitem{StrinNL} D. J. Gross and J. H. Sloan, Nucl. Phys. B \textbf{291},
41 (1987);

W. A. Chemissany, M. de Roo and S. Panda, JHEP \textbf{08}, 037
(2007).

\bibitem{scaling} J. P. Gauntlett, R. C. Myers and P. K. Townsend, Class.
Quantum Gravit. \textbf{16}, 1 (1999).

\bibitem{DEhShakVah} M. H. Dehghani, C. Shakouri and M. H. Vahidinia, Phys.
Rev. D. \textbf{87}, 084013 (2013).

\bibitem{BTZlike} S. H. Hendi, Eur. Phys. J. C \textbf{71}, 1551 (2011).

\bibitem{Gold} N. Goldenfeld, \emph{"Lectures on Phase Transitions and the
Renormalization Group,"} (Westview Press, New York, 1992).

\bibitem{GB} S. W. Wei and Y. X. Liu, Phys. Rev. D \textbf{87}, 044014
(2013);

R. G. Cai, L. M. Cao, L. Li and R. Q. Yang, [arXiv:1306.6233].
\end{thebibliography}
\end{document}